\documentclass[twocolumn,showpacs,preprintnumbers,amsmath,amssymb]{revtex4}
\usepackage{citesort}
\usepackage{epsfig}
\usepackage{graphicx}% Include figure files
\usepackage{dcolumn}% Align table columns on decimal point
\usepackage{bm}% bold math
\usepackage{amsmath}
\newcommand{\be}{\begin{equation}}
\newcommand{\ee}{\end{equation}}
\renewcommand{\vec}[1]{{\bf #1}}
\newcommand{\unitv}[1]{\hat{\bf #1}}
\newcommand{\B}{\vec{B}}

\begin{document}

\preprint{APS/123-QED}

\title{Limits on Weak Magnetic Confinement of Neutral Atoms}
\author{C. A. Sackett}%
\affiliation{Physics Department, University of Virginia,
Charlottesville, VA 22904}
\email{sackett@virginia.edu}
\date{\today}

% ----------------------------------------------------------------
\begin{abstract}
It is shown that when a magnetic field is used to support neutral
atoms against the gravitational force $mg$, 
the total curvature of the field magnitude $B$
must be larger than $m^2 g^2/(2 \mu^2 B)$, where $\mu$ is the
magnetic moment of the atoms. 
This limits the minimum confinement strength obtainable for
a trapped atomic gas.  It is also conjectured that
the curvature must be larger than twice this value for a magnetic
potential that varies in only one or two dimensions, such as an atomic
waveguide.

\end{abstract}

\pacs{03.75.Be,41.20.Gz,39.20.+q}

\maketitle

Magnetic trapping of neutral atoms has become a key technique for the production
and study of Bose-Einstein condensation and other ultra-cold
atomic phenomena.  It is therefore
not surprising that considerable effort has
gone into designing trap configurations with various 
advantages \cite{Inguscio99,Balykin00}.
In most cases, the
trap is used to support the atoms against gravity and to maintain
a density high enough for evaporative cooling.  For this reason, much
attention is paid to making magnetic traps that are as tight and efficient as
possible.  
However, in some situations, tight confinement is not desired.
For instance, by adiabatically expanding
a condensate into a weak trap, extremely low temperatures can be 
obtained \cite{Leanhardt03b,Reeves05}.
Such low temperatures can be useful, for instance, for studying
quantum reflection \cite{Pasquini04} or for probing
atomic interactions very near a Feshbach resonance \cite{Landau77,Tiesinga93}.
The low density achievable in a weak trap may also be useful for
atom interferometry and other applications where 
atomic interactions are undesireable \cite{Shin04,Saba05,Olshanii05}.  
It might be supposed that making a weak trap would be comparatively easy,
but this paper shows that a minimum amount of confinement is
required if the atoms are to remain supported against gravity.

The total potential energy of an atom at position $\vec{x}$
in a gravitational field $\vec{g}$ and
magnetic field $\vec{B}(\vec{x})$ is \cite{Metcalf99}
\be
U(\vec{x}) = mgx_1 + m_F g_F \mu_B |\B|
\ee
where the $x_1$ axis is vertical.  Here $m$ is the atomic
mass, $m_F$ is the magnetic quantum number,
$g_F$ is the Land\'e g-factor for the atomic state, and 
$\mu_B$ is the Bohr magneton. 
For convenience, abbreviate
$\mu = m_F g_F \mu_B$.
Support against gravity evidently requires that the gradient
$\nabla |\B|$ be equal to $-(mg/\mu) \unitv{x}_1$ at the desired trap location.
More generally, suppose a requirement that $\nabla |\B| = \vec{c}$ for
some specified vector $\vec{c}$.

Typically, the potential will have a quadratic dependence near the trapping
location, described by the curvature matrix
\be
K_{ij} = \frac{\partial^2}{\partial x_i \partial x_j}|\B|.
\ee
with eigenvalues $\kappa_i$.  Dependence other than quadratic is
also possible \cite{Bergeman87}.  
Linear dependence, such as that near the zero of
a spherical quadrupole, can be interpreted as the limit
$\kappa_i \rightarrow \infty$ for two or three $\kappa_i$'s.
Higher-order variations of the field may be important as well,
although the results here show that quadratic dependence in at
least one direction is always present if $\vec{c} \neq 0$.

Assuming that the quadratic dependence is most relevant, the
magnetic field will provide three-dimensional
harmonic confinement if $\mu \kappa_i > 0$
for all $i$.  If so, then the atomic oscillation frequencies
are given by
\be
\omega_i^2 = \frac{\mu\kappa_i}{m}.
\ee
Wing has shown that the Laplacian $\nabla^2 |\B|$ 
cannot be negative \cite{Wing84,Ketterle92}.
Since this is the trace of $K$, at least one of the $\kappa_i$ must
be nonnegative, yielding the familiar result that it is not possible to trap
atoms with $\mu < 0$ using a static field.  It is shown here that 
the need for $\nabla|\B|$ to be nonzero provides a more stringent constraint
on $\nabla^2|\B|$ than Wing's theorem alone.

To evaluate $\nabla^2 |\B|$, note
\be
\label{eq-grad}
\begin{split}
\frac{\partial}{\partial x_j} |\B| & = \frac{\partial}{\partial x_j}
\left(\sum_i B_i^2\right)^{1/2}\\
& = \frac{1}{|\B|} \sum_i B_i \frac{\partial B_i}{\partial x_j}.
\end{split}
\ee
Then
\be
\begin{split}
\frac{\partial^2}{\partial x_j^2} |\B| = 
\frac{1}{|\B|} \sum_i \left(\frac{\partial B_i}{\partial x_j}\right)^2
+ \frac{1}{|\B|} \sum_i
B_i \frac{\partial^2 B_i}{\partial x_j^2}\\
- \frac{1}{|\B|^3}\left(
\sum_i B_i \frac{\partial B_i}{\partial x_j}\right)^2.
\end{split}
\ee
Summing over all $j$ yields
\be
\nabla^2 |\B| = \frac{1}{|\B|} \left[
\sum_{ij} \left(\frac{\partial B_i}{\partial x_j}\right)^2
+ \B\cdot(\nabla^2 \B) - \big(
\nabla |\B|\big)^2\right]
\ee
where (\ref{eq-grad}) was used in the last term.  Since the field
satisfies Maxwell's equations, $\nabla^2\B = 0$.  Using
$\nabla |\B| =  \vec{c}$ leaves
\be
\label{eq-laplace1}
\nabla^2 |\B| = \frac{1}{|\B|} \left[
\sum_{ij}\left(\frac{\partial B_i}{\partial x_j}\right)^2
- c^2\right] = 
\frac{1}{|\B|} \left( \sum_{ij} J_{ij}^2 - c^2\right).
\ee
where $c = |\vec{c}|$ and the Jacobian matrix $J$ is defined by
\be
J_{ij} = \frac{\partial B_i}{\partial x_j}.
\ee
Since $\nabla\times\B = 0$, $J$ is symmetric, and since $\nabla\cdot\B = 0$,
the trace of $J$ is zero.  

The Jacobian is also related to the gradient of
$|\B|$.  From (\ref{eq-grad})
\be
\label{eq-matrixmult}
(\nabla |\B|)_j = \sum_i b_i \frac{\partial B_i}{\partial x_j} = 
\sum_i b_i J_{ij} = c_j
\ee
where the vector
$\vec{b} = \B/|\B|$ has magnitude one.  
Since $J$ is real and 
symmetric, it is possible to simplify (\ref{eq-matrixmult})
by working in 
a basis where $J$ is diagonal, with eigenvalues $\lambda_i$.  In this
basis, denoted by a tilde, (\ref{eq-matrixmult}) becomes
\be
\lambda_i \tilde{b}_i = \tilde{c}_i
\ee
for $i = 1$ to 3.
Thus
\be
\label{eq-constraint}
\sum_i \lambda_i^2 \tilde{b}_i^2 = |\tilde{\vec{c}}|^2 = c^2.
\ee
Also in this basis, Eq.~(\ref{eq-laplace1}) becomes
\be
\nabla^2 |\B| = \frac{1}{|\B|} \left(
\sum_{i} \lambda_i^2 - c^2\right).
\ee

It is therefore necessary to determine how small
$\sum \lambda_i^2$ can be, subject to the constraints
(\ref{eq-constraint}), $|\tilde{\vec{b}}| = 1$, and $\sum \lambda_i = 0$.
Using Lagrange multipliers $\alpha$, $\beta$, and $\gamma$, define
\be
F = \sum_i \lambda_i^2 
+ \alpha \sum_i \lambda_i
+ \beta \left(\sum_i \tilde{b}_i^2 - 1\right)
+ \gamma \left(\sum_i\lambda_i^2 \tilde{b}_i^2 - c^2\right).
\ee
Setting to zero the derivatives of $F$ with respect to 
the $\lambda_i$, the $\tilde{b}_i$, and
the multipliers yeilds the constrained extrema of $\sum \lambda_i^2$.
Since it is necessary that at least one of the $\tilde{b}_i$ be nonzero,
take $\tilde{b}_1 \neq 0$.  Then the optimum solution is $\tilde{b}_2 
= \tilde{b}_3 = 0$,
$\lambda_1 = c$, and $\lambda_2 = \lambda_3 = -\lambda_1/2$.
This indicates that generally, 
\be
\sum_i \lambda_i^2 \geq \frac{3}{2} c^2
\ee
so that 
\be
\nabla^2 |\B| \geq \frac{c^2}{2|\B|}.
\ee
This is the primary result of the paper.
In particular, a trap suspending atoms against gravity must satisfy
\be
\label{eq-condition}
\sum_i \omega_i^2 \geq \frac{m g^2}{2\mu|\B|},
\ee
with an analogous statement possible about the strength of the 
anti-trapping potential for atoms with $\mu < 0$.

For an example of a field which reaches this
constraint,  consider
\be
\begin{split}
\B = \left(\frac{c}{2}x - axz\right)\unitv{x}
+ \left(\frac{c}{2}y - ayz\right)\unitv{y}\\
+\left[B_0 - cz + a\left(z^2 - \frac{x^2 + y^2}{2}\right)\right]\unitv{z}.
\end{split}
\ee
Such a field can be generated, for instance, at the center of a pair
of coaxial coils with unequal currents that are
separated by more than the Helmholtz spacing.
To second order, the magnitude of the field is
\be
|\B| = B_0 - cz + az^2 + \frac{1}{2}\left(\frac{c^2}{4B_0} - a\right)
\left(x^2 + y^2\right).
\ee
So if $c = mg/\mu$, atoms will be supported against gravity at the origin.  
The trap frequencies are then
\be
\label{eq-3dfreqs}
\begin{split}
\omega_x^2 & = \omega_y^2 = \frac{\mu}{m}\left(\frac{c^2}{4B_0} - a\right)\\
\omega_z^2 & = \frac{2\mu a}{m}
\end{split}
\ee
and
\be
\sum_i \omega_i^2 = \frac{\mu c^2}{2mB_0} = \frac{mg^2}{2\mu B_0},
\ee
which matches (\ref{eq-condition}).

For experimental comparison, Ref.~\cite{Leanhardt03b}
reports a trap for $^{23}$Na atoms with oscillation
frequencies of $1.81 \pm 0.05$~Hz, $0.65\pm 0.05$~Hz and $1.2 \pm 0.1$~Hz,
yielding $\sum \omega^2 = 201 \pm 12$ s$^{-2}$.  The field magnitude was 17 G
and the magnetic moment was $\mu_B/2$, so that $mg^2/\mu B_0 = 234$~s$^{-2}$.
The reported frequencies thus violate the condition (\ref{eq-condition})
by slightly less than
three standard deviations, perhaps due to an unreported 
uncertainty in the field magnitude.
Nonetheless, the two values are within 15\% of each other.

It should be noted that the sum of the squares of the trap frequencies
is not the only way to measure the trap confinement strength.
Indeed, the atomic density in a harmonic trap 
generally depends on the geometric
mean of the trap frequencies, 
$\bar{\omega} = (\omega_1 \omega_2 \omega_3)^{1/3}$.  If any
one $\omega_i$ is zero, then $\bar{\omega}$ is zero.
As (\ref{eq-3dfreqs}), shows, it is possible to satisfy
(\ref{eq-condition}) while either one or two
frequencies equal zero, for $a = 0$ and $a = c^2/4B_0$ respectively.
However, higher order terms in the potential may still 
be present.  For the above example, when 
$a = c^2/4B_0$ quartic terms provide confinement in $x$ and $y$.
When $a = 0$, cross terms like $x^2 z$ break the translation symmetry
along $z$.
It is therefore also interesting to consider whether (\ref{eq-condition})
still holds for a potential that is truly flat in one or two
dimensions.

A straightforward way to generate true two-dimensional confinement
is to use a field whose components are independent of one coordinate
in some basis.  For instance, the field
\be
\label{eq-2dfield}
\B = (cy - 2ayz)\unitv{y} + \left[B_0 - cz + a(z^2-y^2)\right]\unitv{z}
\ee
can be obtained along a line hanging below two long wires stretched along $x$
and separated along $y$.
It has  a magnitude (to second order) of
\be
|\B| = B_0 - cz + \left(\frac{c^2}{2B_0} - a\right) y^2
+a z^2
\ee
that is manifestly independent of $x$.  In such a case, the Jacobian matrix
$J$ will have, in the appropriate basis, a column consisting
only of zeros, since the derivatives with respect to that coordinate
vanish.  The matrix therefore has one eigenvalue, say $\lambda_3$, that is
zero.  In this case, the minimization of $\sum \lambda^2$ is
straightforward. The trace condition requires $\lambda_1 = -\lambda_2$,
so the gradient condition can be written
\be
\lambda_1^2 \sum_{i=1}^2 b_i^2 = c^2
\ee
and the sum is necessarily less than or equal to one.  Thus
$\lambda_1^2 \geq c^2$ and $\sum \lambda_i^2 \geq 2c^2$.
This requires
\be
\nabla^2|\B| \geq c^2
\ee
or
\be
\label{eq-2dconstraint}
\sum_i \omega_i^2 \geq \frac{m g^2}{\mu|\B|},
\ee
twice the value as the three-dimensional case.  The field
(\ref{eq-2dfield}) satisfies this constraint exactly.

It is possible, however, for the components of the field to vary
along a direction while the magnitude of the field does not.
For instance, the field 
\be
\label{eq-1dfield}
\B = B_0e^{-kz}(\sin ky \unitv{y} + \cos ky \unitv{z})
\ee
can be obtained above a surface magnetized with a sinusoidally oscillating
pattern along $y$ \cite{Vladimirskii61,Roach95}.
Here the field varies with $y$, but the magnitude
\be
|\B| = B_0 e^{-kz}
\ee
does not.  By choosing $k = mg/\mu|\B|$, atoms can be supported against
gravity at any desired position $z$.
Since the field here is still independent of $x$,
the argument of the preceding paragraph applies and it can be seen that
(\ref{eq-1dfield}) exactly satisfies the two-dimensional
constraint (\ref{eq-2dconstraint}).
It may be, however, that an analogous three dimensional field configuration 
exists providing either one- or two-dimensional confinement and
with $\sum \omega^2 < m g^2/\mu|\B|$.  It has proven difficult to 
either find such an example or to show that one does not exist.

In conclusion, it has been demonstrated that a magnetic field
capable of supporting atoms against gravity will necessarily
include a minimum degree of harmonic confinement in at least one direction.
It is notable that the confinement can decrease as the magnitude of the field
itself increases.  This result should prove useful in guiding the 
design of weakly confining traps for studying extremely low-temperature
and low-density gases.

This work was supported by the National Science Foundation and the 
US Office of Naval Research.  The author thanks B. Deissler and 
O. Garcia for their comments on the manuscript.

%\bibliographystyle{apsrev}
%\bibliography{sackett}

\begin{thebibliography}{16}
\expandafter\ifx\csname natexlab\endcsname\relax\def\natexlab#1{#1}\fi
\expandafter\ifx\csname bibnamefont\endcsname\relax
  \def\bibnamefont#1{#1}\fi
\expandafter\ifx\csname bibfnamefont\endcsname\relax
  \def\bibfnamefont#1{#1}\fi
\expandafter\ifx\csname citenamefont\endcsname\relax
  \def\citenamefont#1{#1}\fi
\expandafter\ifx\csname url\endcsname\relax
  \def\url#1{\texttt{#1}}\fi
\expandafter\ifx\csname urlprefix\endcsname\relax\def\urlprefix{URL }\fi
\providecommand{\bibinfo}[2]{#2}
\providecommand{\eprint}[2][]{\url{#2}}

\bibitem[{\citenamefont{Inguscio et~al.}(1999)\citenamefont{Inguscio,
  Stringari, and Wieman}}]{Inguscio99}
\bibinfo{editor}{\bibfnamefont{M.}~\bibnamefont{Inguscio}},
  \bibinfo{editor}{\bibfnamefont{S.}~\bibnamefont{Stringari}},
  \bibnamefont{and} \bibinfo{editor}{\bibfnamefont{C.~E.}
  \bibnamefont{Wieman}}, eds., \emph{\bibinfo{title}{{Bose-Einstein}
  Condensation in Atomic Gases}}, vol. \bibinfo{volume}{CXL} of
  \emph{\bibinfo{series}{Proceedings of the International School of Physics
  {``Enrico Fermi''}}} (\bibinfo{publisher}{IOS Press},
  \bibinfo{address}{Amsterdam}, \bibinfo{year}{1999}).

\bibitem[{\citenamefont{Balykin et~al.}(2000)\citenamefont{Balykin, Minogin,
  and Letokhov}}]{Balykin00}
\bibinfo{author}{\bibfnamefont{V.~I.} \bibnamefont{Balykin}},
  \bibinfo{author}{\bibfnamefont{V.~G.} \bibnamefont{Minogin}},
  \bibnamefont{and} \bibinfo{author}{\bibfnamefont{V.~S.}
  \bibnamefont{Letokhov}}, \bibinfo{journal}{Rep. Prog. Phys.}
  \textbf{\bibinfo{volume}{63}}, \bibinfo{pages}{1429} (\bibinfo{year}{2000}).

\bibitem[{\citenamefont{Leanhardt et~al.}(2003)\citenamefont{Leanhardt,
  Pasquini, Saba, Schirotzek, Shin, Kielpinski, Pritchard, and
  Ketterle}}]{Leanhardt03b}
\bibinfo{author}{\bibfnamefont{A.~E.} \bibnamefont{Leanhardt}},
  \bibinfo{author}{\bibfnamefont{T.~A.} \bibnamefont{Pasquini}},
  \bibinfo{author}{\bibfnamefont{M.}~\bibnamefont{Saba}},
  \bibinfo{author}{\bibfnamefont{A.}~\bibnamefont{Schirotzek}},
  \bibinfo{author}{\bibfnamefont{Y.}~\bibnamefont{Shin}},
  \bibinfo{author}{\bibfnamefont{D.}~\bibnamefont{Kielpinski}},
  \bibinfo{author}{\bibfnamefont{D.~E.} \bibnamefont{Pritchard}},
  \bibnamefont{and} \bibinfo{author}{\bibfnamefont{W.}~\bibnamefont{Ketterle}},
  \bibinfo{journal}{Science} \textbf{\bibinfo{volume}{301}},
  \bibinfo{pages}{1513} (\bibinfo{year}{2003}).

\bibitem[{\citenamefont{Reeves et~al.}(2005)\citenamefont{Reeves, Garcia,
  Deissler, Baranowski, Hughes, and Sackett}}]{Reeves05}
\bibinfo{author}{\bibfnamefont{J.~M.} \bibnamefont{Reeves}},
  \bibinfo{author}{\bibfnamefont{O.}~\bibnamefont{Garcia}},
  \bibinfo{author}{\bibfnamefont{B.}~\bibnamefont{Deissler}},
  \bibinfo{author}{\bibfnamefont{K.~L.} \bibnamefont{Baranowski}},
  \bibinfo{author}{\bibfnamefont{K.~J.} \bibnamefont{Hughes}},
  \bibnamefont{and} \bibinfo{author}{\bibfnamefont{C.~A.}
  \bibnamefont{Sackett}}, \bibinfo{journal}{Phys. Rev. A}
  \textbf{\bibinfo{volume}{72}}, \bibinfo{pages}{051605(R)}
  (\bibinfo{year}{2005}).

\bibitem[{\citenamefont{Pasquini et~al.}(2004)\citenamefont{Pasquini, Shin,
  Sanner, Saba, Schirotzek, Pritchard, and Ketterle}}]{Pasquini04}
\bibinfo{author}{\bibfnamefont{T.~A.} \bibnamefont{Pasquini}},
  \bibinfo{author}{\bibfnamefont{Y.}~\bibnamefont{Shin}},
  \bibinfo{author}{\bibfnamefont{C.}~\bibnamefont{Sanner}},
  \bibinfo{author}{\bibfnamefont{M.}~\bibnamefont{Saba}},
  \bibinfo{author}{\bibfnamefont{A.}~\bibnamefont{Schirotzek}},
  \bibinfo{author}{\bibfnamefont{D.~E.} \bibnamefont{Pritchard}},
  \bibnamefont{and} \bibinfo{author}{\bibfnamefont{W.}~\bibnamefont{Ketterle}},
  \bibinfo{journal}{Phys. Rev. Lett.} \textbf{\bibinfo{volume}{93}},
  \bibinfo{pages}{223201} (\bibinfo{year}{2004}).

\bibitem[{\citenamefont{Landau and Lifshitz}(1977)}]{Landau77}
\bibinfo{author}{\bibfnamefont{L.~D.} \bibnamefont{Landau}} \bibnamefont{and}
  \bibinfo{author}{\bibfnamefont{E.~M.} \bibnamefont{Lifshitz}},
  \emph{\bibinfo{title}{Quantum Mechanics (Non-relativistic Theory)}}
  (\bibinfo{publisher}{Pergamon}, \bibinfo{address}{Oxford},
  \bibinfo{year}{1977}), \bibinfo{edition}{3rd} ed., \bibinfo{note}{{\S} 133}.

\bibitem[{\citenamefont{Tiesinga et~al.}(1993)\citenamefont{Tiesinga, Verhaar,
  and Stoof}}]{Tiesinga93}
\bibinfo{author}{\bibfnamefont{E.}~\bibnamefont{Tiesinga}},
  \bibinfo{author}{\bibfnamefont{B.~J.} \bibnamefont{Verhaar}},
  \bibnamefont{and} \bibinfo{author}{\bibfnamefont{H.~T.~C.}
  \bibnamefont{Stoof}}, \bibinfo{journal}{Phys. Rev. A}
  \textbf{\bibinfo{volume}{47}}, \bibinfo{pages}{4114} (\bibinfo{year}{1993}).

\bibitem[{\citenamefont{Shin et~al.}(2004)\citenamefont{Shin, Saba, Pasquini,
  Ketterle, Pritchard, and Leanhardt}}]{Shin04}
\bibinfo{author}{\bibfnamefont{Y.}~\bibnamefont{Shin}},
  \bibinfo{author}{\bibfnamefont{M.}~\bibnamefont{Saba}},
  \bibinfo{author}{\bibfnamefont{T.~A.} \bibnamefont{Pasquini}},
  \bibinfo{author}{\bibfnamefont{W.}~\bibnamefont{Ketterle}},
  \bibinfo{author}{\bibfnamefont{D.~E.} \bibnamefont{Pritchard}},
  \bibnamefont{and} \bibinfo{author}{\bibfnamefont{A.~E.}
  \bibnamefont{Leanhardt}}, \bibinfo{journal}{Phys. Rev. Lett.}
  \textbf{\bibinfo{volume}{92}}, \bibinfo{pages}{050405}
  (\bibinfo{year}{2004}).

\bibitem[{\citenamefont{Saba et~al.}(2005)\citenamefont{Saba, Pasquini, Sanner,
  Shin, Ketterle, and Pritchard}}]{Saba05}
\bibinfo{author}{\bibfnamefont{M.}~\bibnamefont{Saba}},
  \bibinfo{author}{\bibfnamefont{T.~A.} \bibnamefont{Pasquini}},
  \bibinfo{author}{\bibfnamefont{C.}~\bibnamefont{Sanner}},
  \bibinfo{author}{\bibfnamefont{Y.}~\bibnamefont{Shin}},
  \bibinfo{author}{\bibfnamefont{W.}~\bibnamefont{Ketterle}}, \bibnamefont{and}
  \bibinfo{author}{\bibfnamefont{D.~E.} \bibnamefont{Pritchard}},
  \bibinfo{journal}{Science} \textbf{\bibinfo{volume}{307}},
  \bibinfo{pages}{1945} (\bibinfo{year}{2005}).

\bibitem[{\citenamefont{Olshanii and Dunjko}(2005)}]{Olshanii05}
\bibinfo{author}{\bibfnamefont{M.}~\bibnamefont{Olshanii}} \bibnamefont{and}
  \bibinfo{author}{\bibfnamefont{V.}~\bibnamefont{Dunjko}},
\bibinfo{note}{eprint cond-mat/0505358}.

\bibitem[{\citenamefont{Metcalf and van~der Straten}(1999)}]{Metcalf99}
\bibinfo{author}{\bibfnamefont{H.~J.} \bibnamefont{Metcalf}} \bibnamefont{and}
  \bibinfo{author}{\bibfnamefont{P.}~\bibnamefont{van~der Straten}},
  \emph{\bibinfo{title}{Laser Cooling and Trapping}}
  (\bibinfo{publisher}{Springer}, \bibinfo{address}{New York},
  \bibinfo{year}{1999}).

\bibitem[{\citenamefont{Bergeman et~al.}(1987)\citenamefont{Bergeman, Erez, and
  Metcalf}}]{Bergeman87}
\bibinfo{author}{\bibfnamefont{T.}~\bibnamefont{Bergeman}},
  \bibinfo{author}{\bibfnamefont{G.}~\bibnamefont{Erez}}, \bibnamefont{and}
  \bibinfo{author}{\bibfnamefont{H.~J.}~\bibnamefont{Metcalf}},
  \bibinfo{journal}{Phys. Rev. A} \textbf{\bibinfo{volume}{35}},
  \bibinfo{pages}{1535} (\bibinfo{year}{1987}).

\bibitem[{\citenamefont{Wing}(1984)}]{Wing84}
\bibinfo{author}{\bibfnamefont{W.~H.} \bibnamefont{Wing}},
  \bibinfo{journal}{Prog. Quant. Elect.} \textbf{\bibinfo{volume}{8}},
  \bibinfo{pages}{181} (\bibinfo{year}{1984}).

\bibitem[{\citenamefont{Ketterle and Pritchard}(1992)}]{Ketterle92}
\bibinfo{author}{\bibfnamefont{W.}~\bibnamefont{Ketterle}} \bibnamefont{and}
  \bibinfo{author}{\bibfnamefont{D.~E.} \bibnamefont{Pritchard}},
  \bibinfo{journal}{Appl. Phys. B} \textbf{\bibinfo{volume}{54}},
  \bibinfo{pages}{403} (\bibinfo{year}{1992}).

\bibitem[{\citenamefont{Vladimirski{\^i}}(1961)}]{Vladimirskii61}
\bibinfo{author}{\bibfnamefont{V.~V.} \bibnamefont{Vladimirski{\^i}}},
  \bibinfo{journal}{Sov. Phys. JETP} \textbf{\bibinfo{volume}{12}},
  \bibinfo{pages}{740} (\bibinfo{year}{1961}).

\bibitem[{\citenamefont{Roach et~al.}(1995)\citenamefont{Roach, Abele, Boshier,
  Grossman, Zetie, and Hinds}}]{Roach95}
\bibinfo{author}{\bibfnamefont{T.~M.} \bibnamefont{Roach}},
  \bibinfo{author}{\bibfnamefont{H.}~\bibnamefont{Abele}},
  \bibinfo{author}{\bibfnamefont{M.~G.} \bibnamefont{Boshier}},
  \bibinfo{author}{\bibfnamefont{H.~L.} \bibnamefont{Grossman}},
  \bibinfo{author}{\bibfnamefont{K.~P.} \bibnamefont{Zetie}}, \bibnamefont{and}
  \bibinfo{author}{\bibfnamefont{E.~A.} \bibnamefont{Hinds}},
  \bibinfo{journal}{Phys. Rev. Lett.} \textbf{\bibinfo{volume}{75}},
  \bibinfo{pages}{629} (\bibinfo{year}{1995}).

\end{thebibliography}

\end{document}